\documentclass[iop,aj,twoside,onecolumn,english]{aastex62}

\usepackage{graphicx}
\usepackage{url}
\usepackage{longtable}
\usepackage{amssymb}
\usepackage{multirow}
\usepackage{rotating}

\usepackage{color}
\definecolor{purple}{rgb}{.9,0,.1}
\usepackage{natbib}
\begin{document}

\journalinfo{submitted to AJ 20190715; revised 20190925; revised 20191012} 

\newcommand{\kms}{km s$^{-1}$}
\newcommand{\msun}{M$_{\sun}~$}
\newcommand{\rsun}{R$_{\sun}~$}
\newcommand{\lsun}{L$_{\sun}~$}
\newcommand{\goi}{Gaia 19ajj}

\title{Gaia\,19ajj: A Young Star Brightening Due to Enhanced Accretion $+$ Reduced Extinction}

\author{Lynne A. Hillenbrand} 
\affiliation{Department of Astronomy, California Institute of Technology, Pasadena CA 91125}
\email{lah@astro.caltech.edu}

\author[0000-0001-8174-1932]{Bo Reipurth}
\affiliation{Institute for Astronomy, University of Hawaii at Manoa, 
          640 N. Aohoku Place, Hilo, HI 96720, USA}

\author[0000-0002-8293-1428]{Michael Connelley} 
\affiliation{Institute for Astronomy, University of Hawaii at Manoa, 
          640 N. Aohoku Place, Hilo, HI 96720, USA}

\author[0000-0002-0077-2305]{Roc M. Cutri}
\affiliation{IPAC, California Institute of Technology, Pasadena CA 91125}

\author[0000-0002-0531-1073]{Howard Isaacson} 
\affiliation{Astronomy Department, University of California, Berkeley, CA 94720}
\affiliation{University of Southern Queensland, Toowoomba, QLD 4350, Australia}

\begin{abstract}
We report on the source \goi, identifying it as a young star associated with
a little-studied star-forming region seen along a complex line-of-sight through the Gum Nebula.
The optical lightcurve recently recorded by {\it Gaia} exhibits a slow 
and unsteady 5.5 mag rise over about 3 years, while the mid-infrared
lightcurve from {\it NEOWISE} 
over the same time period shows a 1.2 mag rise having similar structure.
Available color information is inconsistent with pure extinction reduction 
as the cause for the photometric brightening.
Optical spectroscopic characteristics in the current bright phase include: 
little in the way of absorption except for the hallmark \ion{Li}{1} 6707 \AA\ 
signature of youth plus weak e.g. \ion{Ca}{1} and notably \ion{Ba}{2}; 
strong wind/outflow in \ion{Ca}{2}, \ion{Mg}{1} b, \ion{Na}{1} D, H$\alpha$, \ion{K}{1}, \ion{O}{1}; 
jet signatures in [\ion{O}{1}], [\ion{S}{2}], [\ion{Ca}{2}], [\ion{Fe}{2}], and [\ion{Ni}{2}]; 
and narrow rest-velocity emission in neutral species such as \ion{Fe}{1}, \ion{Ni}{1}, and \ion{Mg}{1}. 
The infrared spectrum is also characterized by outflow and emission, including: 
a hot \ion{He}{1} wind, jet lines such as [\ion{Fe}{2}] and $H_2$;
and weak narrow rest-velocity atomic line emission. 
The $^{12}CO$ bandheads are weakly in emission, but there is also broad $H_2O$ absorption.
\goi\ exhibited a previous bright state in the 2010-2012 time frame.
The body of photometric and spectroscopic evidence suggests that the source 
bears resemblance to V2492 Cyg (PTF 10nvg) and PV Cep, both of which 
similarly experience bright phases that recur on long timescales, with
large-amplitude photometric variations and emission-dominated spectra.   
We interpret the behavior of \goi\ as caused by cycles of enhanced disk accretion accompanied by reduced extinction.
\end{abstract}


\section{INTRODUCTION}

The evolution of circumstellar material around young stars 
is a topic of great interest for problems ranging from
the build-up of stellar mass during star formation and early stellar evolution 
to the formation of planets in circumstellar disks.  These processes are heavily influenced over the first 
several Myr of a star's life by the trades between -- on the one hand --
envelope infall and disk accretion bringing mass inward, and -- on the other hand  --
outflows, jets, and winds arising from a range of locations 
in the stellar/circumstellar environment, that eject mass from the system.

One way of tracing dynamical effects in young stars is through photometric
variability.
For nearly a century, highly variable astronomical objects have been reported
in the literature \citep[e.g.][]{joy1945,herbig1946} that were later associated with 
young stars \citep{ambartsumian1949,herbig1952,herbig1957}. The many flavors of 
young star variability were subsequently characterized by
various authors, notably by \cite{herbst1994} and \cite{herbst1999};
see also \cite{ismailov2005}.
Over the past decade, the true diversity of young star behavior 
in the time domain has become more fully appreciated and the lightcurve categories more rigorously defined. 
Increasingly higher cadence and more photometrically precise datasets 
\cite[e.g.][]{cody2014,cody2018}, as well as long-duration, multi-decade 
investigations \cite[e.g.][]{ibryamov2015,mutafov2019} have contributed.
Such studies have been possible due to dedicated monitoring efforts
over small fields, and modern all-hemisphere, and even all-sky, 
time domain surveys. The involvement of amateur astronomers
with sophisticated equipment and eyes on the sky has also been important,
particularly in the identification of rare large-amplitude brightness changes.  

For nearly a half century, we have recognized a small sample of 
large and very large-amplitude ($>3$ mag) young star variables 
as outbursting sources \citep{herbig1977,herbig1989,connelley2018}.
The basic paradigm of episodic accretion, or punctuated periods of enhanced 
mass accretion/outflow that builds up the final $\sim$10\% of the stellar mass,
was developed based on these early but scant observations \citep[see review by][]{HK1996}.
The rates for the different types of outbursts remain 
relatively poorly constrained empirically \citep{hillenbrand2015},
though for recent significant progress see \cite{cp2019}.  
Especially over the past decade, 
our understanding of the diversity in behavior of young stars during 
such large-amplitude brightenings has been enhanced by the coordination 
of multi-color photometric and multi-wavelength spectroscopic follow-up 
of detected brightening events. 

In this paper, we describe a large-amplitude brighting of the newly appreciated young stellar object \goi.
The {\it Gaia} mission, although primarily an astrometric mission, 
offers public alerts\footnote{Gaia Alerts;\ {\url{http://gsaweb.ast.cam.ac.uk/alerts}}}
based on photometric changes exceeding 2 mag in the broad-band optical G filter  
\citep[][Hodgkin et al. 2020, in preparation]{hodgkin2013}.
One such alert was issued on 2019, January 31 for \goi.  Since that time,
the source at position 08:10:45.78 -36:04:30.94 (J2000.)
has continued to brighten.

In this paper, we first describe the source \goi\ and its environment, in \S2.  
In \S3, we report on the publicly available {\it Gaia} lightcurve of 
\goi\ and its historical context based on ASAS and VPHaS photometry. 
We also present WISE and NEOWISE photometry at mid-infrared wavelengths
covering essentially the same time baseline.
Our follow-up near-infrared photometry is presented in \S4 and
the critical spectroscopy at both optical and infrared wavelengths in \S5.
\S6 contains a summary and short discussion of the context of this source 
amid the complex zoo of young star variables, and \S7 our conclusions 
regarding its similarity to sources like 
V2492 Cyg \citep[PTF 10nvg;][]{covey2011,hillenbrand2013,giannini2018}
with PV Cep another good analog.

\section{THE SOURCE}

\subsection{Environment}

\begin{figure}
\centering
\includegraphics[width=0.95\textwidth]{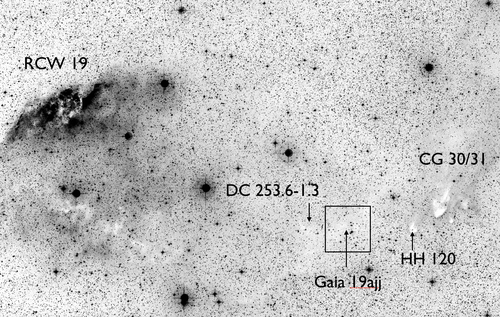} 
\vskip 0.3truein
\includegraphics[width=0.60\textwidth]{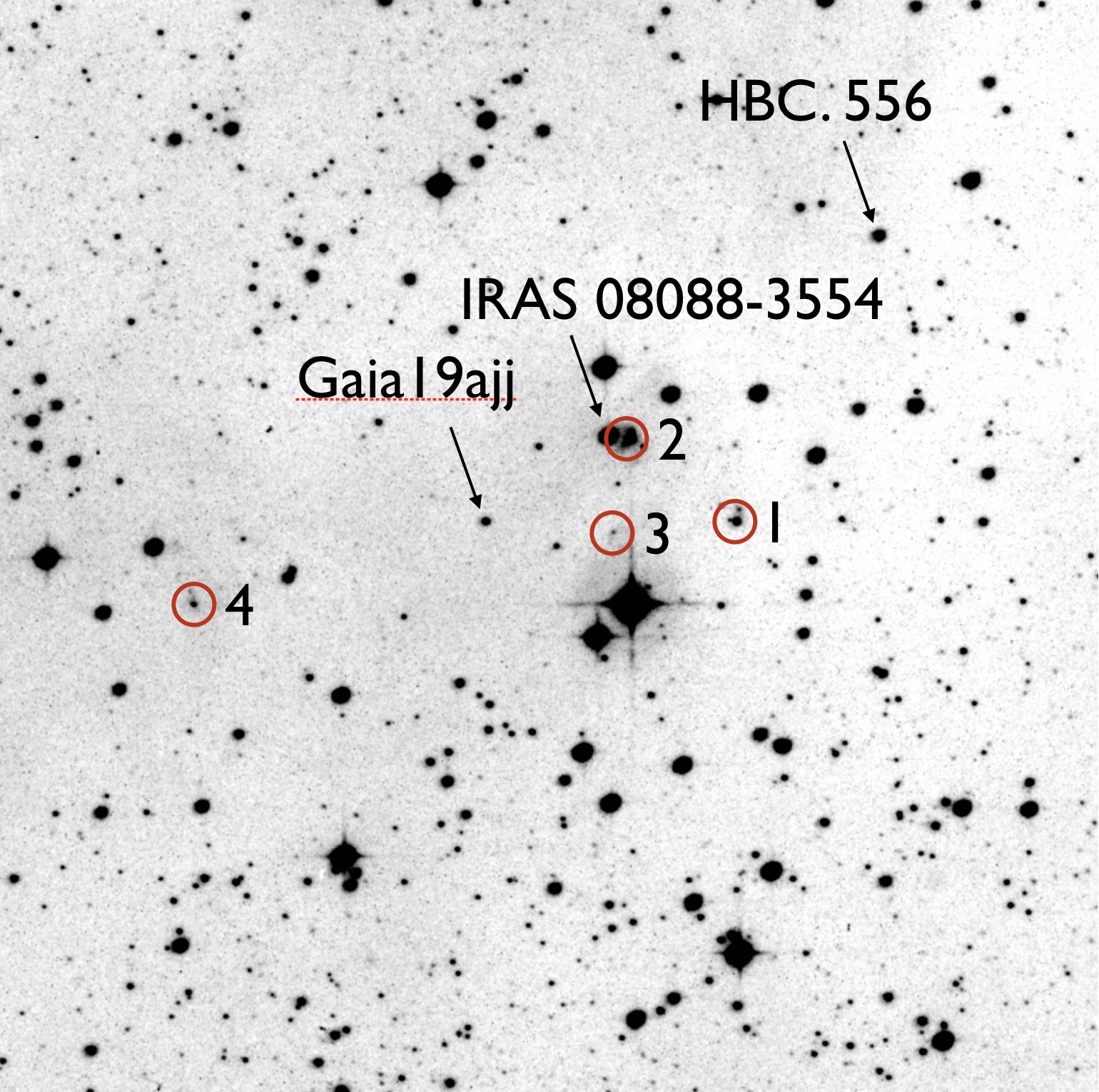} 
\caption{
{\it Top:}
AAO red Schmidt plate from the SES survey
over $2^\circ\times 1^\circ$ showing the general region.
\goi\ is located in a small dark cloud to the southeast of the CG 30/31 complex.
The box indicates the area shown in the bottom panel. 
{\it Bottom:}
UKST SuperCOSMOS H$\alpha$ image over $10\arcmin\times 10\arcmin$
highlighting the location of \goi.
Newly identified compact emission nebulae are numbered as are
various objects discussed in the text.
The image orientation is standard, with north upwards 
and east leftwards, to within 1-2$^\circ$ over the wide field. 
}
\label{fig:region}
\end{figure}

\goi\ is located within an optically opaque small cloud, 
positioned about 1.5 deg below the southern galactic plane. 
Figure~\ref{fig:region} shows the large scale region. 
The sightline passes the northern part of the Vela Molecular Ridge,  
broadly located towards the Gum Nebula.
The best known objects in this general area are the CG~30/31 globules containing HH~120, 
located 15\arcmin\ to the west \citep{reipurth1983}. About a degree to the NE 
is the prominent \ion{H}{2} region RCW~19 = Gum~10 \citep{rodgers1960,gum1955}. 

Between these two features is a small cloud, 
Dobashi 5180 \citep{dobashi2011}, in which \goi\ is located.  
Other very nearby cataloged structures include the dark clouds 
DC~253.6-1.3 \cite{hartley1986} and G253.56-1.32 \citep{db2002}, 
and the Planck-identified dust clump and dense core 
PLCKECC G253.49-01.26 \citep{planck2011}, to the east. 
The \goi\ position is just southeast of IRAS 08088-3554 
which is identified as a compact \ion{H}{2} region.

The immediate region of \goi\ features many diffuse objects in optical images 
from the NOAO DECam Plane Survey \cite[DECaPS;][]{schlafly2018}, 
and there is evidence of a grouping of bright red stars, including \goi, 
in the mid-infrared {\it WISE} images.  
The local cloud is clearly a recently active star-forming region. 
The most well-studied young star in the vicinity is HBC 556, a few arcmin to the west.
\cite{pettersson1987a,pettersson1987b} identified this source as PH$\alpha$21, of spectral type M4, 
in a broad-area search for H$\alpha$ emission stars in the region between the cometary globules and RCW~19.  
In order to gauge the level of star formation activity in this region,
we looked for the presence of additional H$\alpha$ emission sources in the vicinity of \goi\ 
by examining an H$\alpha$ image obtained as part of the AAO SuperCOSMOS survey \citep{parker2005}. 
We identify four such sources in Figure~\ref{fig:region}. 
A color image from DECam (available via the ALADIN service)
shows even more detail of these and additional, fainter nebulae in the vicinity.

\subsection{Distance}
     
The Gaia DR2 \citep{gdr2} distance to \goi, derived by simply inverting 
the measured parallax ($1.200 \pm 0.201$ mas), is $832_{-120}^{+170}$ pc.
This can be compared to previous estimates of distance in this region of sky. 
First, \cite{westerlund1963} discovered an OB association, which he called Puppis~OB~III, 
apparently related to the RCW~19 \ion{H}{2} region to the east of \goi,
and estimated its distance at around 1700~pc. 
Considering the more proximately located HBC 556, 
\cite{pettersson1987a} assumed that this star is located within the Gum Nebula, 
at the same approximate distance of the cometary globules to the west of \goi, 
estimated at around 450 pc. 
We concur that from its brightness and spectral type, HBC 556 can not be a distant object. 
Finally, a kinematic distance of 3.3 kpc 
was suggested by \cite{bhatt1998} for the associated small cloud itself, 
but this value is both highly uncertain and improbable given the other evidence. 
We adopt the 832 pc distance from Gaia DR2.
This implies that \goi, and thus presumably the Dobashi 5180 small cloud, is located behind the Gum Nebula.

\subsection{Spectral Energy Distribution}

\begin{figure}
\includegraphics[width=0.5\textwidth]{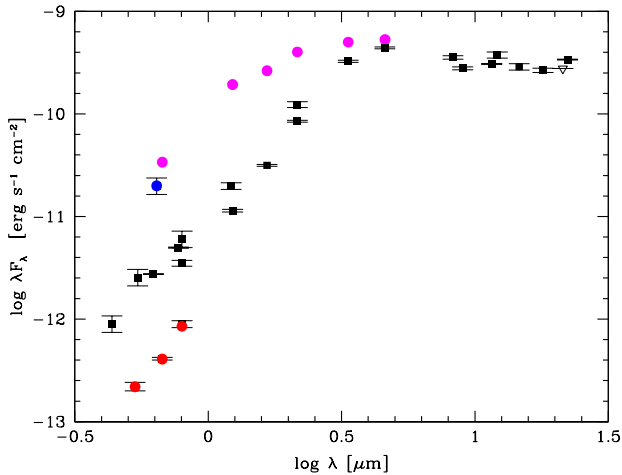}
\vskip-1.5truein
\caption{
Non-simultaneous spectral energy distribution of \goi. 
Historical photometric variability of the source is apparent
in a notably brighter photographic plate measurement (blue) 
and the notably fainter Gaia DR2 magnitudes (red),
relative to the other recorded optical data points
that define a reasonably consistent SED.
Recent data from Gaia, IRTF, and NEOWISE (magenta) 
are brighter and bluer than any previously recorded measurements.
The SED indicates a `Class I' young stellar object type. 
}
\label{fig:sed}
\end{figure}

The spectral energy distribution (SED) that can be assembled for \goi\
(Figure~\ref{fig:sed}; see also the Vizier version\footnote{{\url{http://vizier.u-strasbg.fr/vizier/sed/?submitSimbad=Search\&-c=08:10:45.78\%2B-36:04:30.94\&-c.r=2\&-c.u=arcsec}}}) 
rises from 0.5 to 4 $\mu$m, then turns over with flat 
or only slightly declining energy out to 24 $\mu$m. 
The SED is comprised of photometry taken at different epochs 
-- when the source was in various photometric states. 
Catalog data originates from the NOMAD \citep{zacharias2005}, USNO-B \citep{monet2003}, 
VPHaS DR2 \citep{drew2014}, Gaia DR2 \citep{gdr2}, 2MASS \citep{cutri2003,skrutskie2006}, 
DENIS\footnote{http://cdsweb.u-strasbg.fr/denis.html},
WISE \citep{wright2010}, MSX \citep{price2001}, and AKARI \citep{ishihara2010} catalogs. 
The $G, BP, RP$ photometry from Gaia DR2
is an order of magnitude fainter than the somewhat earlier VPHaS optical photometry 
and much fainter than the decades earlier USNO-B value (see Figure~\ref{fig:lc}).

Regardless of the demonstrated short-wavelength photometric variability,
\goi\ is clearly a `Class I' type young stellar object
with a flat mid-infrared SED\footnote{
Quantitatively, we measure a spectral index from the 2MASS $K_s$ band
and WISE 22 $\mu$m band photometry, as $+$0.36.}. 
This SED category is associated with disk$+$envelope circumstellar geometry,
suggesting a significant amount of high latitude material above the disk
that is distributed in a more spherical-like geometry, with
ongoing infall and rapid accretion onto the central stellar object.
We do not attempt to model the SED given the 
demonstrated variability and the geometric complications 
for the radiative transfer.

The SED also implies that at least some of the short wavelength light 
may be scattered rather than direct.  This is consistent with the fact that the
fainter-state photometry represented in the Gaia DR2 data does not demand
significantly higher reddening relative to the much brighter state of the 
VPHaS measurements.  As discussed below, the photometric changes in \goi\ are likely not 
explained by extinction changes alone.

\subsection{Other Available Information}

There are no previous publications studying this specific star (2MASS 08104579-3604310), 
though it does appear based on its {\it WISE} colors
in the large catalog of \cite{marton2016} identifying candidate young stellar objects. 

\goi\ is an apparent H$\alpha$ emitter, even in previously fainter states.
The VPHaS catalog \citep{drew2014} provides a brightness $r=16.86$ mag  
and a moderate $r-i$ color of $1.13 \pm 0.02$ mag, along with an emission index
$r-H\alpha=0.89 \pm 0.02$ mag. 
According to the models in Figure 2 of \cite{barentsen2013},
the source would be an obvious H$\alpha$ emitter, with equivalent width
of several tens of Angstroms.

Given the spectral energy distribution, as well as the large amplitude 
photometric variability described below, it is unwise to attempt a luminosity calculation.  
However, the observed optical colors suggest a spectral type earlier than $\sim M3$, 
assuming there is also some color contribution from reddening.
Adopting the $\approx 830$ pc distance, the absolute magnitude 
in an average part of the lightcurve (see below) 
is $M_r = 7.26 - A_r$, where $A_r$ is the unknown value of the extinction 
in the r-band.   A typical luminosity around $1$ \lsun for a young 
pre-main sequence star would suggest $A_r \approx$ 3 mag.

\section{THE LIGHTCURVE}

\subsection{Optical}

\begin{figure}
\includegraphics[angle=-90,width=1.00\textwidth]{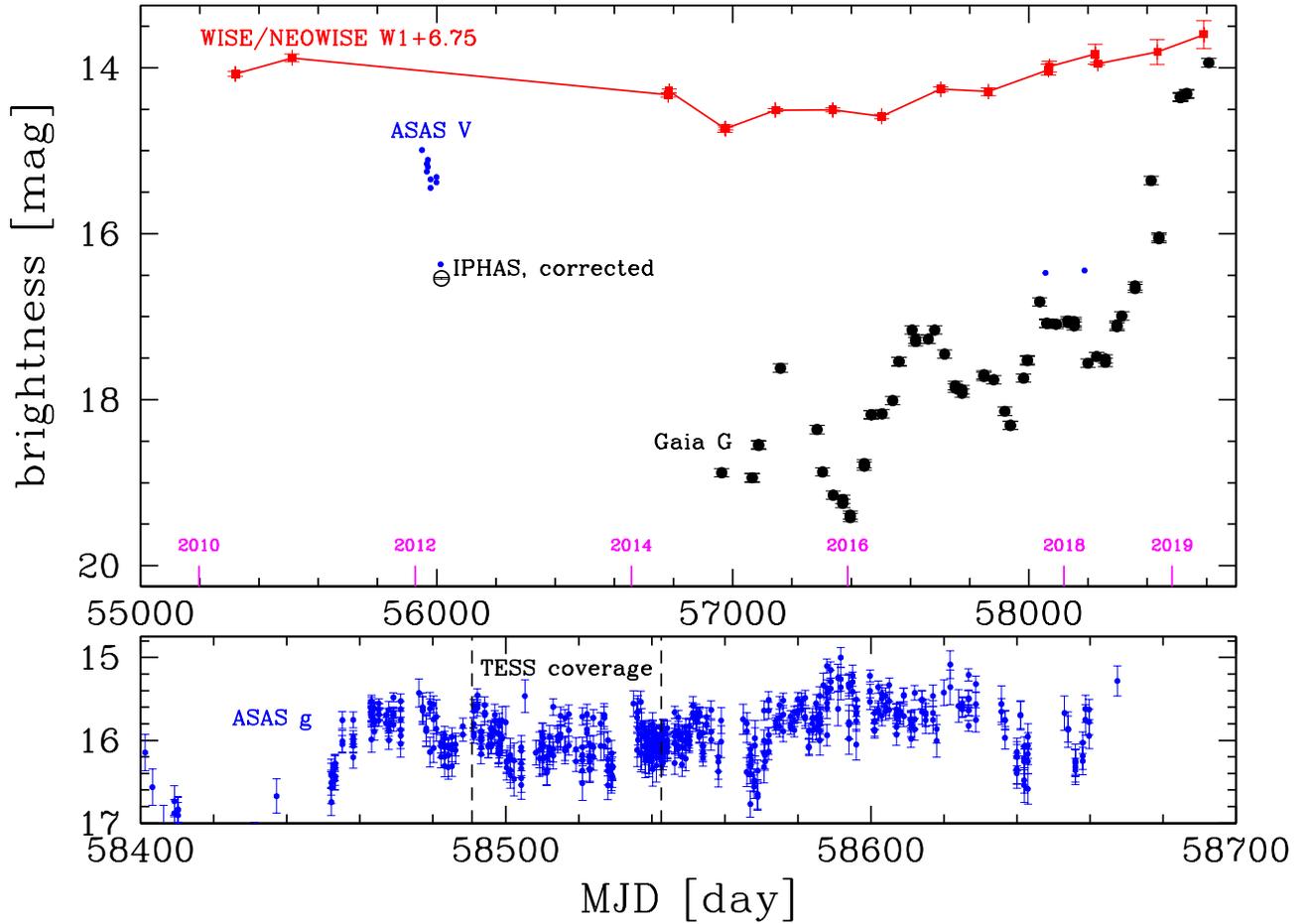}
\caption{
Top panel:
Available lightcurve data for Gaia 19ajj between 2010 and 2019.
Black symbols are {\it Gaia} G-band (effective wavelength 0.673 $\mu$m) 
with assumed 0.05 mag error bars, plus an {\it iPHaS/VPHaS} r-band measurement
converted to G-band.  
Blue points are from ASAS and measured in V-band 
(g-band data points are shown in the bottom panel).
Gaia 19ajj has brightened optically by over 5 mag during the past three years,
but was also bright in the 2010-2012 time frame.
Red points are WISE and NEOWISE measurements at 3.4 $\mu$m, 
shifted downward by 6.75 mag in order to compare to the optical.
The mid-infrared also shows recent long-term brightening, 
exceeding 1 mag over the past three years.
Bottom panel:
Zoom-in to the latest epochs of ASAS data, all in the g-band. 
Short timescale variability accompanies a plateau phase of the recent overall brightening episode;
the oscillations occur on a time scale of several weeks. 
Vertical lines indicate the epochs of field coverage by TESS at high cadence (see text).
}
\label{fig:lc}
\end{figure}

Figure~\ref{fig:lc} shows the {\it Gaia} G-band lightcurve available from
\url{http://gsaweb.ast.cam.ac.uk/alerts/alert/Gaia19ajj/},
with the last update for our analysis occurring on 28 February, 2019.
No error estimates are provided but we adopt 0.05 mag uncertainty.  We include
an {\it iPHaS/VPHaS} \citep{drew2014} r-band measurement that has been
approximately converted to the {\it Gaia} photometric system \citep[using the procedure described
in][]{hillenbrand2018}.  Also shown is the V-band and g-band data
available from {\it ASAS} \citep{shappee2014}.  These data have higher cadence
than the {\it Gaia} data, and indicate quasi-periodic oscillations 
during the 2018-2019 rise, with a periodogram peak around 24 days.
We investigated the availability of TESS data for this source,
which was observed from 7 January through 28 February, 2019, but the TESS
pixels are too big given the field density for a viable lightcurve to be
derived at the current level of TESS data processing.

The total brightening of \goi\ over the past three years exceeds 5.5 mag in the {\it Gaia} G-band.
While impressive, and possibly suggestive of an outbursting source such as an FU Ori star,
\goi\ was also bright in early 2012, conceivably as bright as it is now.
The subsequent large-amplitude fade and more recent re-brightening 
at both infrared and optical wavelengths
suggests that the photometric excursions may have a complex
interpretation, such as being due in part to accretion variations and in part to extinction variations.
A long-duration color time series would help in the assessment.
Unfortunately, there is little optical color information available at present.
By matching the time series between {\it Gaia} and {\it ASAS}, we derive
only a crude $g-G$ color estimate in the range 1.2 - 1.8 mag and no
meaningful color variability information.

The historical record of photographic plate measurements of \goi\ 
supports repeated large-amplitude brightening and fading.
The USNO-A2 \citep{monet1998} records R=17.25 mag for a mean 
epoch between blue and red observations of 1978.5, while the 
USNO-B1 \citep{monet2003} gives R=14.66 mag for mean epoch 1982.3.
In other words, the source has exhibited approximately 3 mag variability
on few year time scales in the past, similar to what has occurred 
over the past 4-7 years.

\subsection{Infrared}

\begin{figure}
\includegraphics[width=0.85\textwidth]{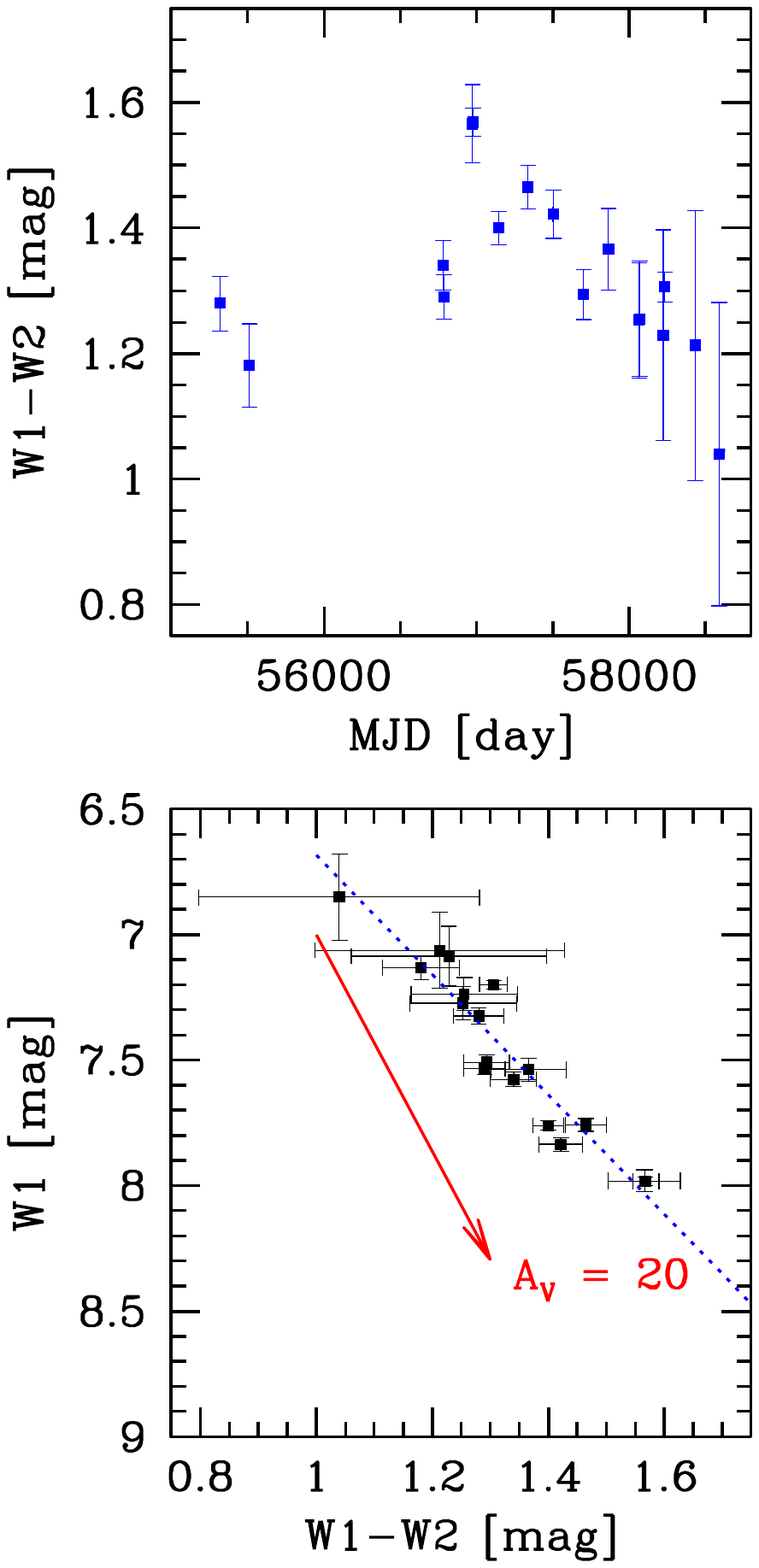}
\vskip-1truein
\caption{
Top panel:
WISE/NEOWISE color evolution between 2010 and 2018 showing
the blueing of the source during the photometric rise illustrated
in Figure~\ref{fig:lc}.  Note that the source was also relatively
blue in 2010, albeit still red in an absolute sense,
which we ascribe as due to circumstellar dust.
Bottom panel:
Color-magnitude diagram in WISE/NEOWISE filters.
A linear fit to the data (blue line) 
demonstrates the distinction of the color-magnitude behavior of \goi\ 
from the expectations for extinction and reddening (red vector).
}
\label{fig:color}
\end{figure}

Variability is also evident in mid-infrared lightcurves.  
\goi\ is well-detected in data from both the primary WISE mission \citep{wright2010}
and the NEOWISE Reactivation mission \citep{mainzer2014}.  

The 3.4 and 4.6 $\mu$m
(W1 and W2) profile-fit photometry for \goi\ was drawn from the 
WISE All-Sky, 3-Band Cryo, Post-Cryo and NEOWISE Reactivation Single-exposure 
Source Databases.  Measurements were used only from exposures for which 
the image quality parameter {\em qual\_frame} was greater than zero, and
on which the source lies  more than 50 pixels from the image edge.  
\goi\ is formally saturated in all WISE/NEOWISE images, requiring additional
attention in order to present quality photometry.  Specifically,  
the profile-fit photometry of saturated sources exhibits a flux-dependent 
bias of varying degrees.  
Therefore, we derived and applied corrections, now published\footnote{Section II.1.c.iv.a of the Explanatory Supplement 
to the NEOWISE Data Release Products \citep{cutri2015}.}, 
to the Source Database magnitudes in order to compensate for the photometric bias.  

The median values of the corrected W1 and W2 magnitudes for each 
observation epoch were computed.  
The uncertainties associated with each median magnitude are the standard deviation
of the population of the corrected magnitudes within each epoch. 
These values are typically slightly smaller than the median of the uncertainties (error)
of the individual corrected magnitudes that include the contribution of
the systematic uncertainties of the photometric bias corrections.
{\it WISE/NEOWISE} photometry is provided in Table 2.

\begin{table}[h]
\begin{centering}
\caption{WISE and NEOWISE Measurements of \textit{Gaia} 19ajj} 
\begin{tabular}{cccccccc}
\hline
MJD& $\sigma$(MJD) & W1 & $\sigma$(W1) & error(W1) & W2 & $\sigma$(W2) & error(W2)\\ 
day& day & mag & mag & mag & mag & mag & mag \\
\hline
55320.08 &  0.404 &  7.324 &  0.031 &  0.051 &  6.044 &  0.025 &  0.066 \\
55511.16 &  0.363 &  7.131 &  0.047 &  0.086 &  5.950 &  0.057 &  0.107 \\
56782.13 &  0.351 &  7.577 &  0.028 &  0.058 &  6.237 &  0.073 &  0.101 \\
56785.19 &  0.316 &  7.533 &  0.025 &  0.062 &  6.243 &  0.055 &  0.102 \\
56973.14 &  0.361 &  7.981 &  0.044 &  0.036 &  6.415 &  0.028 &  0.084 \\
56976.18 &  0.279 &  7.982 &  0.016 &  0.036 &  6.414 &  0.037 &  0.083 \\
57143.97 &  0.387 &  7.760 &  0.019 &  0.043 &  6.360 &  0.044 &  0.085 \\
57337.97 &  0.347 &  7.759 &  0.025 &  0.044 &  6.294 &  0.047 &  0.097 \\
57502.94 &  0.301 &  7.836 &  0.027 &  0.041 &  6.414 &  0.045 &  0.083 \\
57702.24 &  0.376 &  7.508 &  0.028 &  0.065 &  6.214 &  0.068 &  0.104 \\
57863.50 &  0.342 &  7.538 &  0.046 &  0.062 &  6.172 &  0.092 &  0.107 \\
58066.31 &  0.343 &  7.274 &  0.065 &  0.096 &  6.021 &  0.118 &  0.117 \\
58069.14 &  0.365 &  7.237 &  0.065 &  0.103 &  5.982 &  0.093 &  0.117 \\
58223.95 &  0.327 &  7.086 &  0.119 &  0.127 &  5.857 &  0.122 &  0.126 \\
58232.40 &  0.016 &  7.200 &  0.017 &  0.110 &  5.894 &  0.005 &  0.122 \\
58433.38 &  0.329 &  7.062 &  0.152 &  0.127 &  5.849 &  0.117 &  0.126 \\
58590.92 &  0.313 &  6.851 &  0.171 &  0.127 &  5.811 &  0.111 &  0.131 \\
\hline
\end{tabular}
\label{tab:wise}
\end{centering}
\tablecomments{MJD = mean Modified Julian Date.\\
W1,W2 = median WISE 3.4 and 4.6 $\mu$m bias-corrected photometric measurements.\\
$\sigma$ = dispersion in values.\\
error = median of bias-corrected asymmetric uncertainties.}
\end{table}

As illustrated in Figure~\ref{fig:lc}, 
the {\it WISE} data points from 2010 are relatively bright, while
{\it NEOWISE} measured fading between 2014 May and November, 
then gradual brightening from about MJD$\approx$57000
to the present. There is superposed variability 
that generally traces the pattern of the optical fluctuations, 
but at lower amplitude. The cumulative brightening
since 2014 November is about 1.2 mag in W1 (3.6 $\mu$m) compared to 
$>5$ mag in G (0.67 $\mu$m).  

Figure~\ref{fig:color} shows the {\it WISE/NEOWISE} color curve 
which indicates that the source has been getting gradually bluer 
since about MJD$\approx$57000, while it has become brighter (Figure~\ref{fig:lc}). 
Figure~\ref{fig:color} also shows that there is a direct and tight correlation 
between color 
and magnitude.
However, a least-squares fit
indicates that the magnitude-color slope is relatively shallow at
$W1 = 2.39\times(W1-W2) +4.29$ with rms=0.10 mag.
This can be compared to the slope of 4.31 expected for interstellar extinction 
\citep[][adopting the Spitzer extinction law as approximately correct for the WISE bands]{indebetouw2005}.
If interpreted as extinction, the total color change would correspond to
$A_V = \Delta(W1-W2)/(0.56 - 0.43) \times 8.8 = 67.7\times\Delta(W1-W2) \approx 35$ mag.    
However, there is more color change than expected given the magnitude change, which suggests
that at least some of the blueing behavior is intrinsic to the source and not just extinction clearing.

This shallow color slope is also apparent in the near-infrared.
As discussed below, comparing
the changes in $J-H$ vs $H-K$ between 2MASS and new
IRTF photometry shows larger $H-K$ change relative to $J-H$ than 
would be expected under standard interstellar reddening assumptions.
As noted above, the SED also shows an inconsistency between the large change 
in optical brightness relative to only a small change in optical color. 

\subsection{Lightcurve Assessment}
Both the {\it Gaia} lightcurve and the {\it NEOWISE} lightcurve exhibited 
quasi-periodic behavior over the few years prior to
the recent steep optical rise. Then, starting in early 2018 and continuing into 2019, 
the slope changed and \goi\ brightened by another several magnitudes
($\sim$3.5 in the optical and $\sim$1 in the infrared).
The optical slope was -0.31 mag/month over about 10 months.   %

The optical and mid-infrared lightcurve evidence 
for wavelength-dependent, long term photometric brightening 
that is punctuated by brief fading events, suggests that extinction variability
may be an important consideration in interpreting the behavior of \goi.  
However, extinction variations are not the entire story.
Accretion variability is driving some of the brightening and blueing, 
which also has the effect of reducing the amount of dust, and thereby
diminishing the line-of-sight extinction.  This is the same interpretation 
as is invoked for several better-studied sources with well-characterized 
large-amplitude and long-timescale lightcurves, like V2492 Cyg and PV Cep
(see discussion in \S6). 

\section{NEW INFRARED PHOTOMETRY}

\goi\ was observed in JHKL' bands on the night of 3 February, 2019 (UT) using the slit viewing camera of SpeX \citep{rayner2003} at the NASA Infrared Telescop Facility. Conditions were photometric and the source was observed at an airmass of 1.85 resulting in FWHM=0.92", which is sampled at 0.12'' pixel$^{-1}$.  The filters in SpeX are in the MKO photometric system (Simons \& Tokunaga 2002).  We observed UKIRT faint standard FS 123 for the JHK flux calibration, and HD 84800 for the L' flux calibration.  For each image, we subtracted off a dark then divided by a normalized sky flat.  We then subtracted the median sky value from each image, and aligned and coadded the frames together.  Aperture photometry was done with {\it imexam} in IRAF using a target aperture radius of 20, 23, and 26 pixels, a buffer of 10 pixels, and a sky annulus 10 pixels wide.  Since the non-destructive reads and coadds are added together in IRTF data, we were careful to divide by the value of the DIVISOR header keyword.  The resulting magnitudes were corrected 
for airmass using the extinction coefficients for Maunakea in Krisciunas et al. (1987).  

The resulting measurements are J = 10.749; H =  9.634; K =  8.406; L' = 7.135. 
The formal errors are about 0.01 mag, but the real errors are dominated by the systematics
of the airmass correction, and thus larger than this.

Transformation to the 2MASS photometric system can be achieved using 
the equations in \cite{connelley2007}.
Relative to the JHK photometry reported in the 2MASS catalog \citep{cutri2003},
the recently measured colors of \goi\ are bluer in a manner that might indicate
motion roughly along a reddening vector, 
with  a reduction in $A_V$ of about 6 mag suggested.   
However, there is more $H-K$ difference relative to the $J-H$ difference 
than expected.  Some of the blueing color could be caused by accretion
effects rather than reduction in extinction, similar to our interpretation
of the WISE/NEOWISE color changes.

\section{NEW INFRARED AND OPTICAL SPECTROSCOPY}
\subsection{Observations}

We obtained near-infrared spectra using the IRTF and SpeX \citep{rayner2003} 
on 3 February 2019 (UT) in both SXD (0.7-2.4 $\mu$m) and LXD (1.7-4.2 $\mu$m) modes.
The resolution with the 0.5" slit was $R\approx 1200$.  For the SXD data 
the realized S/N per exposure is $\sim$80 in the Y-band and 
$\sim$300 in the K-band (with 10 exposures taken for 30 minutes
of total exposure time).  For the LXD data the S/N is $\sim$50 per exposure 
in L' (with 16 exposures taken for 8 minutes of total exposure time).  
The K-band seeing was about 0.5" at zenith but closer to 1" for observations 
taken over 1.78 to 1.95 airmasses at the target position.  
A second SXD spectrum was obtained on 26 March 2019 (UT) 
under somewhat worse seeing conditions.
All data were reduced with SpeXtool \citep{cushing2004}.  
In addition to the spectroscopy, a small set of $H_2$ and broad-band $K$
images were obtained on the 26 March observing date, 
and show spatially extended $H_2$ emanating from the southeast
(despite the poor seeing of $\sim 1.9\arcsec$), extending to about 7$\arcsec$.

We obtained an optical echelle spectrum between $\sim$3400-7900 \AA\ at resolution 
R$\approx$60,000 using the Keck I telescope and HIRES \citep{vogt1994} 
on 14 February, 2019 (UT).
Data acquisition used the standard operating procedures of the 
California Planet Search described in \cite{howard2010}.  
Under temporarily clear skies and $\sim$2.0" seeing, a 1500 sec exposure 
resulted in a spectrum with S/N = 15 at 5500 \AA. 
We aligned the \ion{Li}{1} absorption feature to determine
a source radial velocity of $+35.8$ \kms.

Finally, we obtained an infrared echelle spectrum in the 1 $\mu$m Y-band region
at R$\approx$37,000 with the Keck II telescope and the recently upgraded 
\citep{martin2018} NIRSPEC \citep{mclean1998} on 8 April 2019 (UT).  
Two rounds of A-B-B-A position nods were taken with individual 
exposure times of 150 sec per nod position.  The seeing was 0.5-0.6" overhead,
though the target was observed at 1.8 airmasses.
These data were processed using the REDSPEC package\footnote{written by L. Prato, S.S. Kim, \& I.S. McLean} and the resulting S/N = 60 at 1.08 $\mu$m.

\begin{figure}
\includegraphics[width=1.00\textwidth]{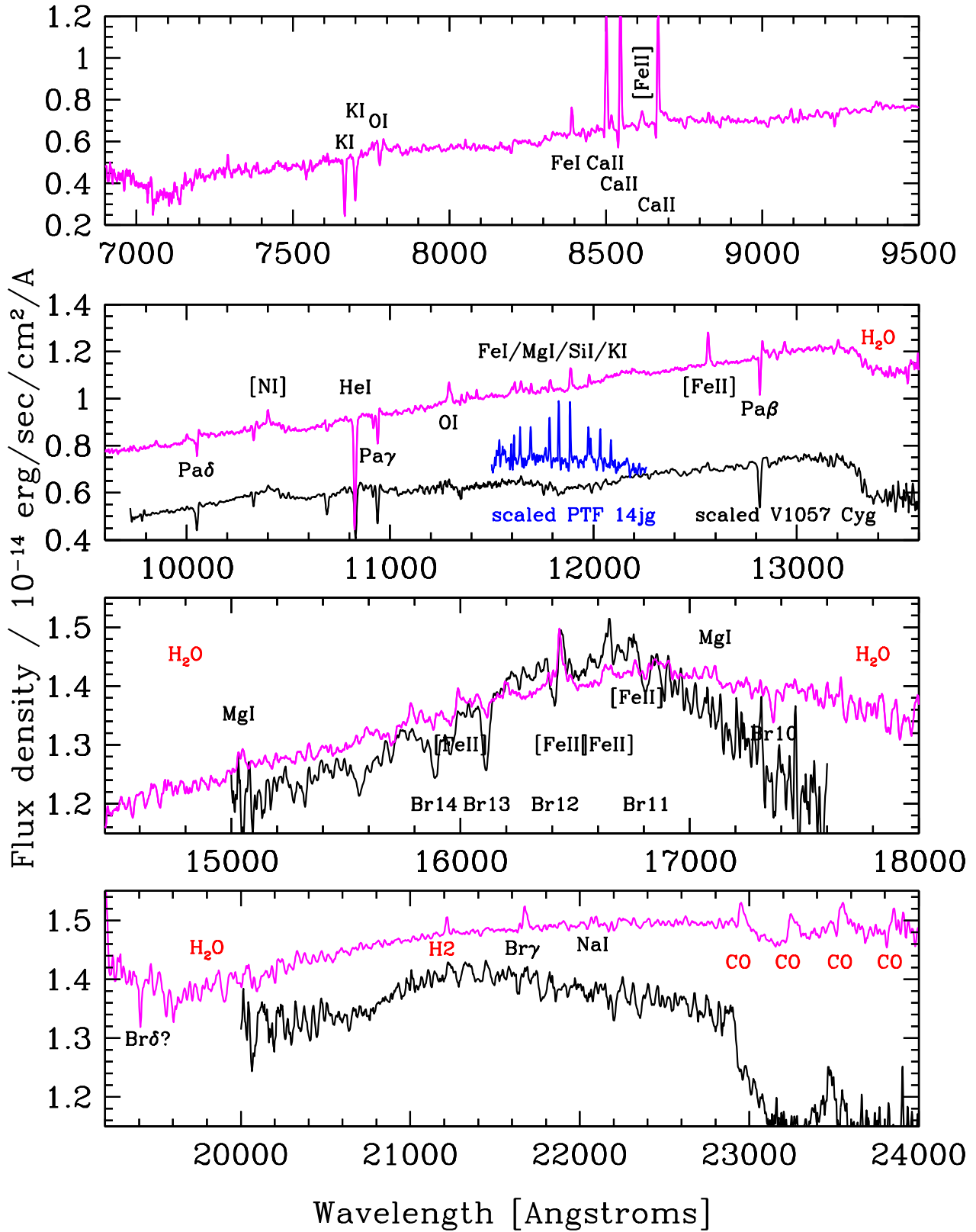}
\caption{
IRTF/SpeX spectrum of \goi\ in magenta
compared with a Palomar/TripleSpec spectrum of V1057 Cyg in black,
showing similarity in the overall spectral shape and broad absorption,
and with a snippet of the spectrum of PTF 14jg in blue, showing similar but stronger
emission in the J-band.
Various permitted and forbidden atomic lines are labelled (black), as are 
molecular lines and broad bands (red).
}
\label{fig:irspec}
\end{figure}

\begin{figure}
\includegraphics[width=1.00\textwidth]{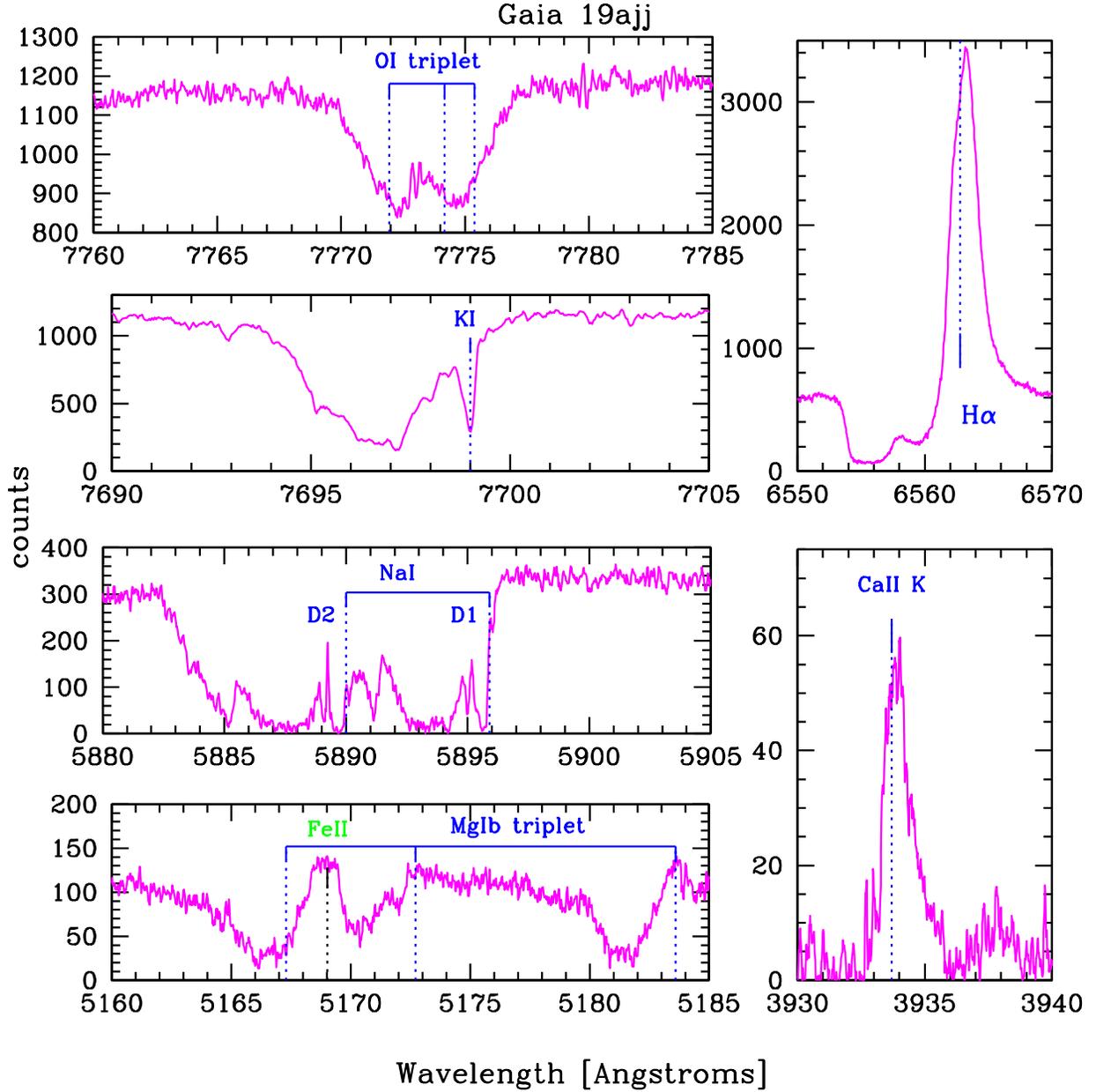}
\vskip -1.5truein
\caption{
Portions of the Keck/HIRES spectrum illustrating the strong wind of \goi\
via blueshifted absorption features.  
The H$\alpha$ profile has a clear P-Cygni nature,
extending to about $-450$ km/s on the blue side, with the red peak at $+25$ km/s.
The \ion{Ca}{2} K line may have a similar P-Cygni profile, but the signal is low 
near the continuum level.  Blueshifted absorption is also seen in  
the \ion{Mg}{1} b triplet lines (with the bluest line contaminated by blueshifted \ion{Fe}{2} 5169 \AA), 
the \ion{Na}{1} D doublet lines, the \ion{K}{1} doublet, and the \ion{O}{1} triplet - all signatures of a strong wind.  
}
\label{fig:wind}
\end{figure}

\begin{figure}
\includegraphics[width=1.00\textwidth]{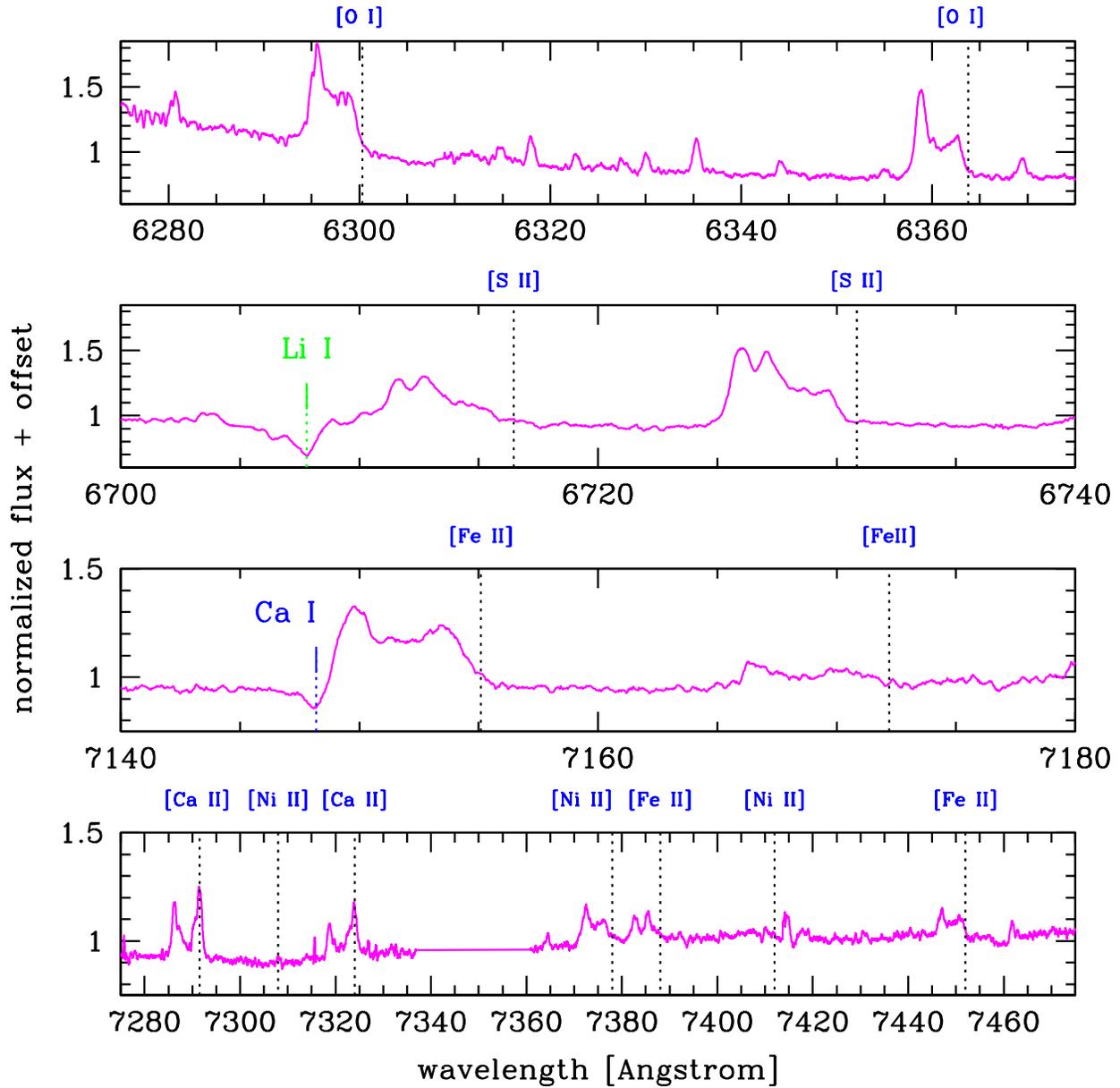}
\vskip -1.5truein
\caption{
Portions of the Keck/HIRES spectrum illustrating the blueshifted forbidden emission lines, 
indicative of the presence of shocked gas in an outflowing jet.
Note that the wavelength range differs among the panels.
All species have multiply peaked profiles, with only the [\ion{Ca}{2}] lines 
having an emission component at the rest velocity.
\goi\ also exhibits clear \ion{Li}{1} 6707 \AA\ absorption.
}
\label{fig:jet}
\end{figure}

\begin{figure}
\includegraphics[width=0.80\textwidth,angle=-90]{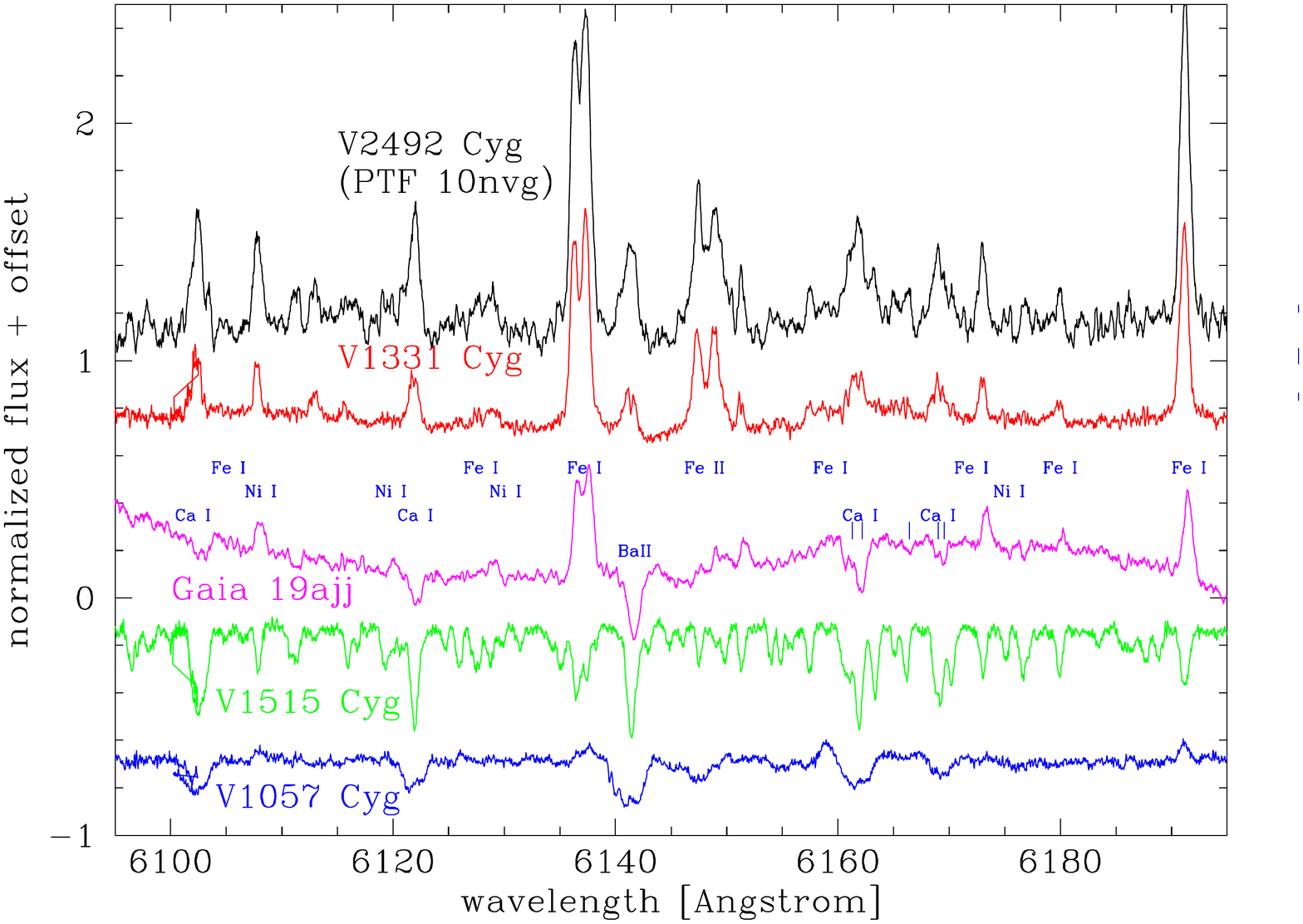}
\caption{
Portion of the HIRES spectrum highlighting the narrow emission lines 
and the weak absorption component of \goi. 
The emission series is similar to, but less extreme than the spectra exhibited
by V2492 Cyg in its bright state (an object with a rather similar, large-amplitude lightcurve) 
and by V1331 Cyg, which is a well-known accretion-dominated object 
with a stochastic but relatively constant lightcurve (varying within 0.5 mag) over time.
The absorption-dominated FU Ori stars V1515 Cyg and V1057 Cyg are also shown for comparison. 
There is little rest-velocity absorption in \goi, with this order unusually rich
in such lines (labelled \ion{Ca}{1} and \ion{Ba}{2}, all of which are also present in the FU Ori stars). 
}
\label{fig:emission}
\end{figure}

\subsection{Spectrum Characterization}

Figures~\ref{fig:irspec}-\ref{fig:nirspec} illustrate some details of our spectra
and are discussed in the subsections below,  with the main features as follows.
The \goi\ optical and infrared spectra exhibit many prominent lines that are indicative
of outflowing material. There is evidence for both a neutral stellar/disk wind 
absorbing continuum radiation, and also shocked gas indicative of an
outflowing jet.  Weak metallic emission lines appear at the rest velocity.

In the optical, notably absent is strong photospheric absorption,
with a few weak metallic lines identified, as discussed below. 
Several absorption features can be attributed to diffuse interstellar bands.
In the infrared, the SXD observations in the YJHK bands indicate a red continuum
with broad absorption due to molecular species.
The LXD observations in L-band exhibit a flat/blue continuum shape. 
\subsubsection{Wind Lines}

\goi\ exhibits classic P Cygni structure in the Balmer H$\alpha$ feature,
with redshifted emission peaking around $+25$ \kms\ and blueshifted absorption  
extending to $-450$ \kms\ (Figure~\ref{fig:wind}).  Similar structure 
is seen more weakly in the infrared in the Pa$\beta$, Pa$\gamma$, and Pa$\delta$ lines,
albeit from much lower spectral resolution (Figure~\ref{fig:irspec}). 
At Br$\gamma$ there is simple emission, and Br$\delta$ and Br10 
are in absorption, with very weak absorption or no sign of higher Br lines.

There is also a P Cygni signature in the \ion{Ca}{2} ``infrared" triplet lines around 8500 \AA\
(Figure~\ref{fig:irspec}) that is likely present as well in the \ion{Ca}{2} K line at 3933 \AA\  (though masked by low S/N; Figure~\ref{fig:wind}).
Other lines with wind signatures in the form of purely blue-shifted absorption include: 
\ion{Fe}{2} 5018, 5169, and 5316, the \ion{Mg}{1} b triplet, the \ion{Na}{1} D doublet,
the \ion{K}{1} 7665/7699 doublet, the \ion{O}{1} 7774 triplet (Figure~\ref{fig:wind}), 
and the higher excitation \ion{He}{1} 10830 triplet 
(Figures~\ref{fig:irspec} and ~\ref{fig:nirspec}). 

The hot \ion{He}{1} line is very strong, reaching 50\% 
of the continuum level in the earlier SpeX spectrum and all the way to 5\% of the
continuum in the later NIRSPEC spectrum, with the terminal velocity 
around $-450$ \kms.  This assumes that the rest velocity is at a heliocentric
velocity of $+35.8$ \kms\ which was the optical correction based on the
\ion{Li}{1} line, and indeed matches the location of the photospheric  \ion{Mg}{2} $+$ \ion{Sr}{2} line in the 1$\mu$m spectrum.
There is no signature of a feature in either absorption or emission at \ion{He}{1} 5876 \AA, 
which is the next transition higher from the 10830 \AA\ upper level, and is optically thin.  
The wind models of \cite{kwan2007} which treat the \ion{He}{1} 10830 line formation 
as a purely scattering process from the continuum photons are consistent with our observations. 
More specifically, the breadth and depth of the \goi\ line profile (Figure~\ref{fig:nirspec}) 
seem to match the genre of \cite{kwan2007} models that adopt a {\it stellar} wind geometry 
rather than an inner disk wind geometry.  Empirically, the \ion{He}{1} profile 
of \goi\ most resembles that of the strong outflow source SVS 13,
as illustrated in \cite{edwards2006}.

\subsubsection{Jet Lines}

\iffalse

  -- look for [SII] in nirspec spectrum -- means reducing different order
   1.028955 
   1.028955 
   1.032332 
   1.032332 
   1.033924 
   1.033924 
   1.037334 

\fi

Shocked outflowing gas is evidenced by blueshifted and broad emission in the optical doublets of 
[\ion{O}{1}], [\ion{S}{2}], [\ion{Ca}{2}], [\ion{Fe}{2}], and [\ion{Ni}{2}]. 
As illustrated in Figure~\ref{fig:jet},
all of these lines are double-peaked with the [\ion{S}{2}] triply peaked.  In most species
the higher velocity component is stronger, but in [\ion{Ca}{2}] the lower velocity 
component is stronger.  
Uniquely for the [\ion{Ca}{2}] doublet lines, the low-velocity peak is at zero velocity  
relative to the star. 

In the infrared (Figure~\ref{fig:irspec}), jet signatures are manifest in
several lines of [\ion{Fe}{2}], along with shocked $H_2$ emission in the lowest energy lines.
The \ion{O}{1} line at 1.13 $\mu$m is somewhat broad, and has been associated with jets in other young sources.

The [\ion{Ca}{2}], [\ion{Ni}{2}], [\ion{Fe}{2}], and [\ion{S}{2}]
lines exhibited in the optical and infrared spectra of \goi\
have intermediate ionization potential, 6.1-10.4 eV. There is 
no evidence of the often-seen lines of [\ion{O}{2}] or [\ion{N}{2}] which have
higher ionization (13.6-14.5 eV), suggesting moderately low temperature. 
The various ratios among same species lines appear consistent with 
intermediate density ($>10^4-6\times10^5$ cm$^{-3}$) -- except for the [\ion{Ca}{2}].

The [\ion{Ca}{2}] lines are high density \citep{ferland1989,hartigan2004,nisini2005}
and are unusual in T Tauri stars, even those showing the forbidden lines 
discussed above. They have been observed, however, in a number of unusual 
large-amplitude photometric variables such as RW Aur, V2492 Cyg (PTF 10nvg), and PTF 14jg \citep{hillenbrand2019},
as well as in the extreme emission-line young star V1331 Cyg.
As mentioned above, the [\ion{Ca}{2}] is the only forbidden species
in \goi\ having a zero-velocity component (Figure~\ref{fig:jet}).
The ratio of the 7291 \AA\ to the 7324 \AA\ is approximately 1.2:1, slightly
lower than the expected 1.5:1 discussed in great detail by \cite{hartigan2004}.  
The [\ion{Ca}{2}] lines are ground state transitions with their upper
level populated either by collisions upward, or by downward transitions 
via the permitted \ion{Ca}{2} triplet lines.
If the [\ion{Ca}{2}] occurs by fluorescence rather than collisions,
this may explain both the different morphology relative to the other 
forbidden lines, and the slightly low doublet ratio. 

\subsubsection{Narrow Emission and Weak \ion{H}{1} Emission Lines}

There is narrow rest-velocity emission in \goi\ in 
permitted metallic lines such as \ion{Fe}{1}, and \ion{Ni}{1} in the optical, as illustrated in Figure~\ref{fig:emission}, 
There is also weak metal emission in the infrared, 
as illustrated in Figure~\ref{fig:irspec}, which shows these same 
species plus other lines such as
\ion{Mg}{1} and \ion{Si}{1} in J-band and H-band, and \ion{Na}{1} in K-band. 
The line strengths are generally only $W_\lambda \gtrsim -1.0$ \AA.
This narrow emission spectrum is similar to that of the pre-peak spectrum 
of the FU Ori candidate outbursting source PTF 14jg \citep{hillenbrand2019},
the large-amplitude accretion$+$extinction source V2492 Cyg (PTF 10nvg) \citep{hillenbrand2013},
as well as that of the photometrically quiet but extreme emission line star V1331 Cyg.  

The infrared spectrum of \goi\ also shows weak emission in the $CO$ bandheads, 
though perhaps superposed on troughs of absorption (see discussion below).  

\begin{figure}
\includegraphics[width=0.50\textwidth]{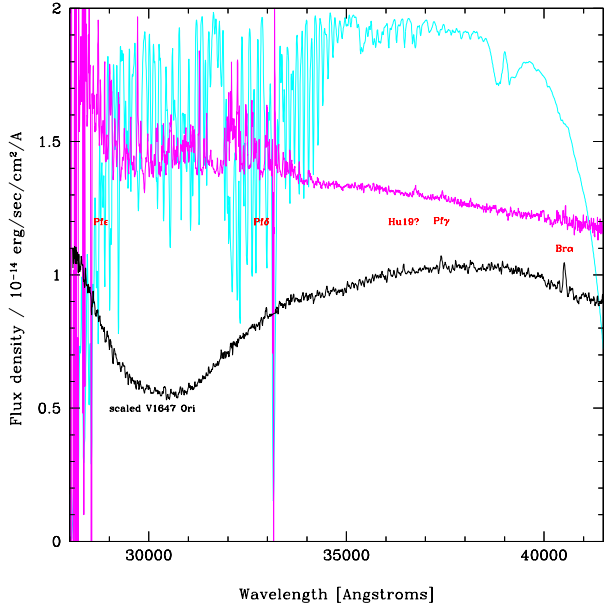}
\vskip-0.5truein
\caption{
The long wavelength spectrum of \goi\ (magenta) compared to the Class I type 
variable V1647 Ori (black), which exhibits prominent broad absorption due to water ice.
In contrast, the \goi\ continuum is slightly blue. 
Note that we attribute the apparent narrow emission in \goi\ 
throughout the water region to inaccurate telluric correction 
(cyan is a scaled atmospheric transmission spectrum).
Weak hydrogen line emission is present in both sources.
}
\label{fig:lxd}
\end{figure}

In the long wavelength spectrum (Figure~\ref{fig:lxd}),  
there is weak \ion{H}{1} emission
as evidenced by Br$\alpha$, Pf$\gamma$ and Pf$\delta$.  
There is another weak line at 3.675 $\mu$m that we can not identify 
(The wavelength seems incorrect for Hu19 at 3.645 $\mu$m 
and there is no corresponding Hu18 at 3.693 $\mu$m).
The spectrum bears some resemblance in showing these weak lines to V1647 Ori
and Z CMa \citep{connelley2018}.

\subsubsection{Interpreting the Strong Emission Lines}

\begin{figure}
\includegraphics[width=0.50\textwidth]{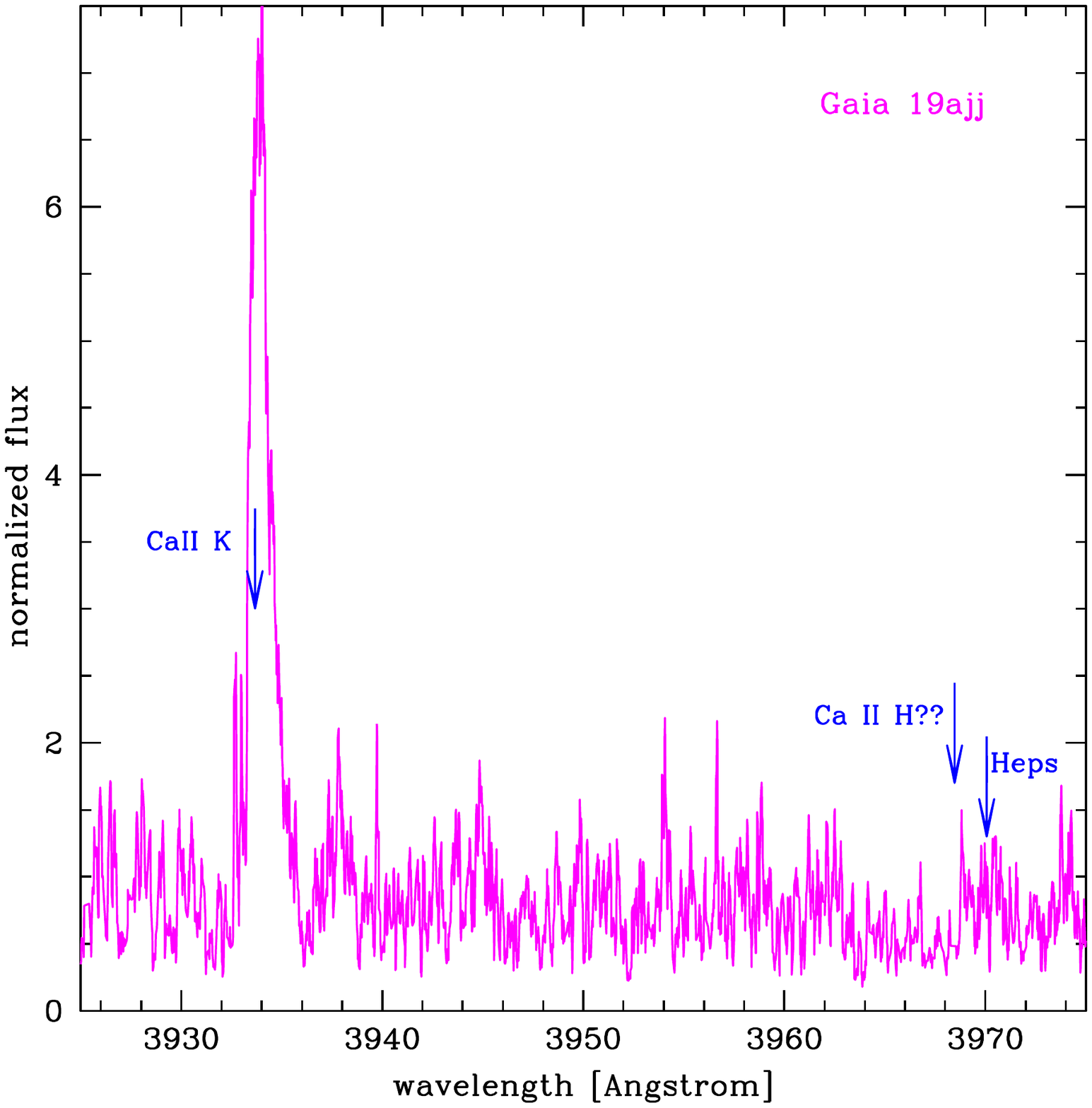}
\vskip-0.5truein
\caption{
The \ion{Ca}{2} H and K spectral region, which is somewhat noisy but
shows strong K yet no signature of H.
These lines are a resonance doublet, and typically exhibit a K:H ratio 
of 1-2 in young stars. The ratio is significantly higher in \goi.
}
\label{fig:hk}
\end{figure}

The strongest emission in \goi\ comes from $H\alpha$ and \ion{Ca}{2},
both associated with the accretion flow in typical young stars when 
these lines are strong and broad.  The $H\alpha$ line has 
$W_\lambda\approx -11$ \AA\ in its redshifted emission component, 
but as discussed above there is also a strong blueshifted absorption 
component ($W_\lambda\approx 5$ \AA) to the overall P-Cygni type profile,
weakening the emission strength.
The 8542 \AA\ line, usually the strongest of the \ion{Ca}{2} triplet lines,
has $W_\lambda=-6.3$ \AA\  
(integrated flux $4\times 10^{-14}$ erg s$^{-1}$ cm$^{-2}$ \AA$^{-1}$),
while the 3933 \AA\ K line, 
usually the stronger of the \ion{Ca}{2} doublet, has $W_\lambda\approx -18$ \AA\
(with large uncertainty due to the low S/N in the continuum in this region).  
Again, there is likely blueshifted absorption in the \ion{Ca}{2} 
lines that reduces the emission strengths.  

We note with some interest the complete absence of the 3968 \AA\ H line
(see Figure~\ref{fig:hk}), which is highly unusual given the generally
comparable fluxes of the two components of the doublet 
(K:H ratio in the range $1-2$) 
in most young accretors \citep[e.g.][]{herczeg2008,rigliaco2012}. 
In active chromospheres the K:H ratio 
increases as the activity level decreases and also occupies the range 
$1-2$ \citep{houdebine1997}.
Suppression of the H line relative to the K line is clear in \goi\
and seems to defy the atomic physics prediction of a maximum ratio of 2
(in the optically thin case from the $g-f$ values).
Despite the doublet ratio oddity, the three lines of the \ion{Ca}{2} triplet, 
which share upper levels with the \ion{Ca}{2} doublet, have 
typical ratios for young star accretors, around $0.95 : 1 : 0.85$
for the $8498 : 8542 : 8662$ \AA\ lines.  The usual interpretation
is that the triplet lines are very optically thick (c.f. the expected 
$0.11 : 1 : 0.55$ proportions for optically thin lines). 

An explanation for the peculiar doublet ratio is that there is 
contaminating P Cygni structure from $H\epsilon$ just 121 km/s redward 
of the \ion{Ca}{2} H line.  However, while $H\alpha$ is strong in \goi, 
the upper Balmer lines are weak with only hints of 
$H\gamma$, $H\delta$, and $H\epsilon$.  A complete cancellation of
\ion{Ca}{2} H emission having a strength $\sim 50-100$\% of the 
\ion{Ca}{2} K line would require a coincidental $H\epsilon$ 
blueshifted absorption component of equal or larger strength, but
without a redshifted emission component that exceeds the very weak 
$H\epsilon$ signal illustrated in Figure~\ref{fig:hk}.
We find that \cite{herbig1989} called attention to the same
peculiarity of a lack of \ion{Ca}{2} H despite strong \ion{Ca}{2} K 
in the pre-outburst spectrum of V1057 Cyg, as well as noting it in
FU Ori in the post-outburst state along with a few other 
unusual young stellar objects.  The H line supression may be a feature
of sources with strong winds having a certain temperature and 
column density or optical depth.

\subsubsection{The Photosphere}

\goi\ is compared in Figure~\ref{fig:irspec} to the FU Ori type star V1057~Cyg,
which shows rather prominent $^{12}CO$ absorption in the 2.3 $\mu$m region
\citep[though weakening recently[]{connelley2018}. 
As noted above, \goi\ shows some signatures of emission 
near the bandhead regions, which turn into a drop in the continuum 
towards redder wavelengths due to the blended CO lines. 
The peculiarity of the $CO$ region aside,
there is resemblance of the \goi\ near-infrared spectrum 
with that of FU Ori stars in the prominent $H_2O$ absorption feature,
which is seen in all of the K-band, H-band, and J-bands \citep{connelley2018}. 
We have two SpeX spectra taken two months apart, and they
show subtle but significant changes in the overall spectrum.
Specifically in this interval, the water vapor absorption bands grew stronger, 
resulting in a more ``triangular" shape through the H-band. 
At the same time, the CO emission lines in the K-band became slightly weaker, 
while the Br$\gamma$ and the H$_2$ emission lines became slightly stronger.
At longer wavelengths, Figure~\ref{fig:lxd} shows only a flat continuum.
There is no broad absorption between 2.8 and 3.3 $\mu$m due to water ice, 
as is seen in many young embedded or Class I type stars \citep[e.g.][]{connelley2018}.  

\begin{figure}
\includegraphics[width=0.50\textwidth]{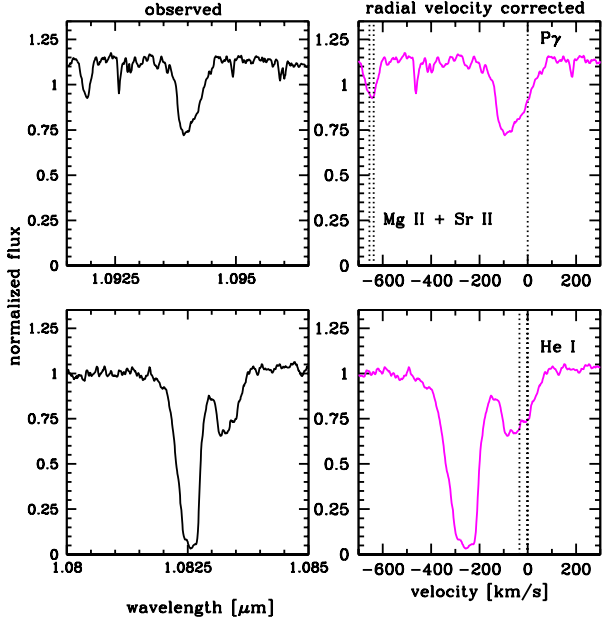}
\vskip-0.5truein
\caption{
Portions of the Keck/NIRSPEC spectrum around \ion{He}{1} 10830 \AA\
and \ion{H}{1} 10938 \AA\ (Pa$\gamma$), illustrating the
blueshifted absorption in these lines (c.f. the lower resolution data 
of Figure~\ref{fig:irspec}).  The narrow features in the Pa$\gamma$ region 
are uncorrected telluric lines.  Left panels are observed. 
Right panels are shifted by the same velocity as applied
to the optical spectrum, resulting in excellent alignment for a narrow
photospheric blend, thus confirming the inferred radial velocity.
The hot \ion{He}{1}
gas is demonstrated to be outflowing with significant occultation 
of the 1 $\mu$m continuum by the outflow region. 
These line profiles can be compared to those exhibited by
the cooler optical wind lines of Figure~\ref{fig:wind}. 
}
\label{fig:nirspec}
\end{figure}

In terms of narrow atomic absorption, the low resolution near-infrared spectrum 
of \goi\ exhibits little of the absorption due to atomic lines 
expected in an FU Ori interpretation. 
Our NIRSPEC spectrum, however, does show a photospheric line (Figure~\ref{fig:nirspec}) that can be attributed to \ion{Sr}{2} and/or \ion{Mg}{2}, 
both intermediate ionization species.
\cite{wallace2000} demonstrate that this feature is prominent in F and G type
supergiants, weaker in giants, and absent in dwarfs.
In the optical, aside from the hallmark \ion{Li}{1} 6707 \AA\ youth signature (Figure~\ref{fig:jet}), 
the strongest rest-velocity absorption that we have identified comes from 
\ion{Ca}{1} and \ion{Ba}{2} (Figure~\ref{fig:emission}).
Another order contains both \ion{Ca}{1} 5857.45 and \ion{Ba}{2} 5853.67.
It is unclear why only \ion{Ca}{1} absorption is strongly present, and not other lines 
having similar excitation.  One clue may be that these lines have some surface
gravity sensitivity for spectral types later than mid-G \citep{prisinzano2012}.
Besides these more obvious lines, 
only hints of weak absorption from other neutral metals appear in the \goi\ high resolution spectrum, 
especially in the 5500-5900 \AA\ region, for example species such as Si, Sc, Mg, and Fe, Ni. 

Regarding \ion{Ba}{2}, the strongest (though blended) \ion{Ba}{2} feature 
at 6497 \AA\ unfortunately is between spectral orders in our HIRES spectrum. 
The illustrated 6142 \AA\ line has $W_\lambda=0.39$ \AA, while
the 5853 \AA\ line has $W_\lambda=0.16$ \AA. 
These equivalent widths are notably stronger than the \ion{Ba}{2} line strengths of
normal dwarf stellar photospheres (of any spectral type). 
These lines are known to have a positive luminosity effect, 
and are strong in supergiants \citep{andrievsky2006}. 
The measured equivalent widths are about 2/3 of the line strengths
seen in FU Ori, V1057 Cyg, and V1515 Cyg.
While the \ion{Ba}{2} line is typically seen in FU Ori sources, including in the
recently announced objects Gaia 17bpi \citep{hillenbrand2018} 
and PTF 14jg \citep{hillenbrand2019}, it is not present in other young stars. 
This is consistent with both the disk atmosphere \citep{HK1985} 
and the distended stellar photosphere \citep{petrov1992} 
interpretation of the optical spectra of FU Ori objects, each of which
are expected to produce low surface gravity spectral signatures.

\section{SUMMARY OF FINDINGS AND DISCUSSION}

The source \goi\ recently exhibited a large-amplitude, long term brightening event,
rising in the optical by more than 5.5 mag over the past three years including 
over 3.5 mag in the past one year.  
The body of evidence suggests that the source is a young star
associated with a little-studied small dark cloud located behind the Gum Nebula.
The source appears to have been comparably bright about 7 years ago.  
Its recent color and magnitude behavior in the time domain suggests that 
variable extinction is part of the explanation for the slow undulative variations 
in the lightcurve during its overall large-amplitude rise. 
However, episodic accretion likely also plays a role since the lightcurve and color-curve data
is inconsistent with pure extinction clearing.  Furthermore, in its current bright state,
the optical and infrared spectra of \goi\ exhibit strong accretion and outflow signatures.
Similar to the early outburst stages of several categories of eruptive young stars, 
the strong wind has a rather high terminal velocity.  

Some of the spectroscopic characteristics of \goi\ are FU Ori-like in nature.
FU Ori stars are accretion outbursts, exhibiting large-amplitude phototometric rises 
over months to years, with decay times of decades to centuries.
Similarities include: mild water absorption in the near-infrared,  
rest-velocity \ion{Ba}{2} and \ion{Li}{1} in the optical, and strongly blueshifted, 
prominent absorption in several metal lines, indicative of a strong wind. 
The terminal wind velocity is higher in \goi\ than in many FU Ori stars,
but objects such as V1735 Cyg, Z CMa, Par 21, and V582 Aur, as well as V1647 Ori (see below) 
do have comparable or even larger velocities.
Absent from the \goi\ spectrum, at least in its present state, are the 
broad atomic photospheric absorption features and the obvious spectral type change 
with wavelength that are typical defining characteristics of FU Ori stars.  
The lightcurve is also unlike that of an FU Ori, 
showing a previous bright phase and substantial structure in the recent photometric rise.  

Some spectrosopic characteristics of \goi\ are more like those of
EX Lup-type accretion bursts. EX Lup stars exhibit distinct moderate-amplitude 
phototometric rises over weeks to months, with decay times of months.
Notably in common is the narrow atomic emission seen in both 
the optical and the infrared 
\citep{kospal2011,sa2012}, with the \goi\ 
narrow emission somewhat weaker than generally seen in EX Lup stars.
EX Lup sources lack the strong wind signatures 
(clear P Cygni profiles with sub-continuum blueshifted absorption) 
of the FU Ori stars and exhibited by \goi.

Finally, some of \goi's spectroscopic characteristics are peculiar for either of 
the above outburst classes, notably the strong forbidden-line emission.

We thus look beyond these popular categories of young star brightening and 
develop an analogy of \goi\ to sources that have been interpreted as
accretion-brightening combined with a reduction in the line-of-sight extinction
along the path through the circumstellar envelope to the central young accretion system.

There are many well-known young stars that experience long-duration fades and
recoveries relative to a fairly steady photometric baseline.  Examples include
GM Cep \citep[e.g.][]{huang2019}, V350 Cep \citep[e.g.][]{semkov2017}, and RW Aur \citep[e.g.][]{koutoulaki2019}, 
several of which exhibit UX Ori type behavior (photometric dimming events that are accompanied by reddening 
which turns blueward near the depths of the fades). 
The photometric excursions of these objects are lower amplitude than 
seen in the lightcurve of \goi, and they do not seem to exhibit signs of 
enhanced accretion during the bright phases\footnote{ Although this has been claimed photometrically for GM Cep, the spectroscopy of \cite{giannini2018rnaas} indicates otherwise.}.
A star with a lightcurve more like that of \goi\ is HBC 340 \citep{dahm2017}, 
which also exhibits an undulating photometric rise but, 
like the sources mentioned above, does not have an accompanying enhancement 
in the spectroscopic accretion/wind indicators.

V2492 Cyg has been characterized as a star with large-amplitude magnitude 
and color variations, a manifestation of extinction variations, 
but perhaps ultimately driven by accretion variations
\citep[see e.g.][]{covey2011,kospal2013,hillenbrand2013,giannini2018}. 
A notable aspect of this source, is that in its faint state, 
the spectrum of V2492 Cyg is dominated
by forbidden line emission formed in the jet/outflow 
(which becomes stronger than H$\alpha$ as the source fades). 
The bright state, by contrast, features mostly lines that are 
formed in the magnetospheric infall region or the accretion shock.
\goi's bright state spectrum is not as strong or quite as rich 
in emission lines as V2492 Cyg, but there are definite spectral similarities
(e.g. Figure~\ref{fig:emission}).  However, V2492 Cyg does not exhibit 
the molecular absorption in $H_2O$ nor the possible $CO$ region
absorption that \goi\ shows (e.g. Figure~\ref{fig:irspec}).
We also note that there is no \ion{Ba}{2} signature in V2492 Cyg.

PV Cep is another source with large-amplitude photometric variations
and also an emission-line dominated spectrum with no or very weak absorption in the optical.
Over the past decade, this source experienced a $\approx$3 magnitude unsteady rise over
about 5 years\footnote{See e.g. {\url{https://www.aavso.org}}}, 
quite similar in timescale to \goi\ though not as large in
amplitude, and similar in amplitude to V2492 Cyg but with a shorter timescale.
Before this, it had experienced a flux decrease of $\approx$4 mag accompanied 
by a weakening of the robust \ion{Ca}{2} triplet emission accretion indicator \citep{kun2011}.
The behavior was attributed to a combination of reduced accretion and enhanced extinction effects,
the latter possibly related to changes in the inner disk structure.
\cite{lorenzetti2011} discusses the optical and infrared photometric 
variations of PV Cep.
\cite{caratti2013} discuss the infrared emission line spectrum, which is much stronger than
that of \goi, including clear CO bandhead emission and stronger jet lines (e.g. [\ion{Fe}{2}] and $H_2$).
Despite the stronger emission, the wind signatures of PV Cep appear not as prevalent as they are in \goi.  
It remains an interesting analog, however.

Comparison can also be made to V1647 Ori, which is a repeating burst source
that is sometimes associated with the EX Lup category, but has 
larger amplitude, longer duration, and less frequent bursts than most EX Lup stars;
it is perhaps best characterized as being the prototype of the V1647 Ori category.  
As for V2492 Cyg, while there are large color changes in V1647 Ori, 
extinction variations are not the dominant effect on the lightcurve \citep[e.g.][]{mcgehee2004,ap2007,aspin2009}.  
Further, the permitted emission lines in this source are even stronger than in V2492 Cyg or \goi, 
and there was only weak forbidden emission during the V1647 Ori outbursts.
There is also no \ion{Ba}{2} signature in V1647 Ori.

\section{CONCLUSION}

While it is often tempting for authors to categorize large-amplitude photometric
variability in young stars in terms of known families, the variability phase space 
is still rather incompletely mapped.  On timescales of weeks to months, 
there are clear differences between the variability types exhibited by
Class I sources relative to those in later stages of circumstellar evolution 
(Class II and Class III sources), with Class I variability tending to be 
larger amplitude and longer timescale \citep{rebull2015,wolk2018}.  
Over months to years, both the Class I and the Class II sources can show 
long duration trends, as well as distinct episodic brightening events 
that last months to decades.

\goi\ is an example of a Class I object that is clearly a rapid accretor 
and driving a strong outflow.  The recent photometric brightening is indeed likely
driven in large part by a positive change in the accretion/outflow activity, 
which has also caused a reduction in the line-of-sight extinction to the central star and
inner accretion zone.  Although \goi\ exhibits some FU Ori-like absorption characteristics 
in its spectrum, notably \ion{Ba}{2} in the optical, 
as well as several other low-gravity atomic lines,  
and broad molecular absorption in the near-infrared,
it also displays EX Lup-like emission (in both the optical and the near-infrared J-band).
We conclude that the best analogs may be V2492 Cyg, PV Cep, and possibly V1647 Ori, 
in which there are substantial changes in extinction that accompany
enhanced accretion. All of these objects appear to have a repeating 
brightening and fading pattern on time scales of several years.
It is not yet obvious to which, if any, of the above 
variable-accretion categories \goi\ should be associated.

\section{ACKNOWLEDGMENTS}
We gratefully acknowledge Simon Hodgkin for his stewardship of the 
Gaia Photometric Science Alerts, as well as
the Gaia Photometric Science Alerts Team, DPAC, and ESA/Gaia. 
We also gratefully acknowledge the NASA/NEOWISE Team. 
Andrew Howard enabled the Keck/HIRES data acquisition.
{Trevor David investigated for us the potential availability of TESS data.} 
This research has made extensive use of CDS services such as SIMBAD and Vizier.
This research has benefited from consultation of the AAVSO Database;
we acknowledge with thanks the variable star observations contributed there
by observers worldwide.

\facility{Gaia, ASAS-SN, IRTF:SpeX, Keck:I(HIRES), Keck:II(NIRSPEC), IPHaS/VPHAS, 2MASS, WISE, NEOWISE, IRSA}

\end{document}